\documentclass[prb,twocolumn,showpacs,superscriptaddress]{revtex4-1}

\usepackage{graphicx}
\usepackage{verbatim}
\usepackage{comment}
\usepackage{amssymb}
\usepackage{stmaryrd}

\begin{document}

\title{Numerical study of the thermodynamics of clinoatacamite}
\author{Ehsan Khatami}
\affiliation{Department of Physics, Georgetown University, Washington, DC 20057, USA}
\author{Joel S. Helton}
\affiliation{NIST Center for Neutron Research, National Institute of Standards
and Technology, Gaithersburg, Maryland 20899, USA}
\author{Marcos Rigol}
\affiliation{Department of Physics, Georgetown University, Washington, DC 20057, USA}


\begin{abstract}
We study the thermodynamic properties of the clinoatacamite compound, Cu$_{2}$(OH)$_{3}$Cl,
by considering several approximate models. They include the Heisenberg model on (i) the
uniform pyrochlore lattice, (ii) a very anisotropic pyrochlore lattice, and (iii) a 
kagome lattice weakly coupled to spins that sit on a triangular lattice. We utilize 
the exact diagonalization of small clusters with periodic boundary conditions and implement a 
numerical linked-cluster expansion approach for quantum lattice models with reduced 
symmetries, which allows us to solve model (iii) in the thermodynamic limit. We 
find a very good agreement between the experimental uniform susceptibility and the 
numerical results for models (ii) and (iii), which suggests a weak ferromagnetic 
coupling between the kagome and triangular layers in clinoatacamite. We also study 
thermodynamic properties in a geometrical transition between a planar pyrochlore 
lattice and the kagome lattice.
\end{abstract}

\maketitle

%%%%%%%%%%%%%%%%%%%%%%%%%%%%%%%%%%%%%%%%%%%%%%%%%%%%%%%%%%%%%%%%%%%%%%%%%%%%%%%%%
%                                Introduction                                   %
%%%%%%%%%%%%%%%%%%%%%%%%%%%%%%%%%%%%%%%%%%%%%%%%%%%%%%%%%%%%%%%%%%%%%%%%%%%%%%%%%

\section{Introduction}

The kagome and pyrochlore lattices are among the archetype systems for
highly frustrated magnetism, with both lattices displaying corner-sharing
frustrated plaquettes (triangles for the two-dimensional kagome
lattice and tetrahedra for the three-dimensional pyrochlore
lattice).  There is also a geometric connection between the two lattices, as the
pyrochlore lattice is composed of alternating kagome and triangular lattice
planes stacked on top of each other (along the $\langle 1~1~1 \rangle$ body
diagonal in typical cubic spinels that display a pyrochlore lattice).  This
leads to the possibility of structural pyrochlore lattices where magnetic
interactions differ within kagome planes and between the kagome and
triangular planes.

Several three-dimensional (3D) pyrochlore lattice materials have been shown to 
decouple into kagome planes ordered antiferromagnetically\cite{A_Wills_99,Wiebe_05} 
or ferromagnetically\cite{Matsuda_08,Ross_09} that are fairly well isolated from
the neighboring triangular plane spins. The Zn-paratacamite mineral family, 
Zn$_{x}$Cu$_{4-x}$(OH)$_{6}$Cl$_{2}$, with $x \, \geq$ 0.3 features spin-$\frac{1}{2}$ 
Cu$^{2+}$ ions arranged on an antiferromagnetically coupled kagome lattice 
alternating with triangular lattice layers occupied by either Cu or nonmagnetic Zn ions.  
The $x$~=~1 end member of this family, herbertsmithite, has attracted interest as a 
strong candidate to display a spin-liquid ground state on almost perfectly decoupled 
two-dimensional (2D) kagome layers.\cite{Helton_07,Mendels_07,M_rigol_07c} 
However, the best available samples are likely not stoichiometric,\cite{Freedman_10} 
with a small fraction of Cu ions on the triangular lattice planes weakly (of 
the order of 1~K) coupled to the kagome planes.\cite{Bert_07} Materials 
such as YBaCo$_{4}$O$_{7}$ (Ref.~\onlinecite{Chapon_06}) and 
Y$_{0.5}$Ca$_{0.5}$BaCo$_{4}$O$_{7}$ (Ref.~\onlinecite{Stewart_11}) 
also feature alternating kagome and triangular layers, but with a stacking that 
is structurally distinct from the pyrochlore lattice.

Here, we are interested in the properties of the mineral clinoatacamite,~\cite{j_grice_96} 
a monoclinic polymorph of Cu$_{2}$(OH)$_{3}$Cl that crystallizes in the P2$_{1}$/n 
space group and features spin-$\frac{1}{2}$ Cu$^{2+}$ ions decorated on a distorted 
pyrochlore lattice.  The mineral is the extension of the Zn-paratacamite family to 
$x$~=~0, with the monoclinic distortion that occurs for $x \, <$~0.3.  
Clinoatacamite has drawn attention in recent 
years,~\cite{X_Zheng_05a,X_Zheng_05b,S_Lee_07,A_Wills_08,A_Wills_09,H_Morodomi_10}
in part due to its unique pyrochlore structure and in part due to the still
unexplained nature of successive phase transitions. Some studies\cite{S_Lee_07,H_Morodomi_10} 
have described the lattice as consisting of distorted kagome layers coupled weakly 
through triangular layers of out-of-plane spins. Others have suggested a  
pyrochlore structure with significant couplings of all Cu spins.~\cite{X_Zheng_05a,A_Wills_09} 
Susceptibility and specific-heat measurements display two transitions upon 
cooling, at $T_{c2}$~=~18~K and
$T_{c1} \, \approx$~6.4~K.  Long-range magnetic order\cite{A_Wills_08,Kim_08}
and a weak ferromagnetic moment are present below $T_{c1}$.  For temperatures
$T_{c1} \, < T \, < T_{c2}$, muon oscillations are observed\cite{X_Zheng_05b}
suggesting a static local moment, which was originally attributed to N\'{e}el
order, while neutron diffraction experiments find no sign of ordering in this
temperature range, and the specific heat anomaly at $T_{c2}$ is too small for the
entropy change expected at an ordering transition. Further analysis of this 
unusual phase between 6.4 and 18~K would be aided by a complete knowledge of 
the local bond strengths in this distorted lattice.

In this work, we study the thermodynamic properties of the clinoatacamite compound by 
considering, as approximate descriptions, the antiferromagnetic Heisenberg model on
(i) a uniform pyrochlore lattice, (ii) a very anisotropic pyrochlore lattice,
which can be seen as a quasi-two-dimensional model, and (iii) a kagome 
lattice with weak ferromagnetic coupling to (otherwise disconnected) spins 
sitting on a triangular lattice, i.e., a two-dimensional model. 
We calculate the spin susceptibility, specific heat, and entropy for these models 
using the exact diagonalization (ED) of small clusters with periodic boundary conditions and, 
only for model (iii), by means of an implementation of the numerical 
linked-cluster expansions (NLCEs)~\cite{M_rigol_06,M_rigol_07a} on an anisotropic 
checkerboard lattice that displays the required geometry. NLCEs yield exact results 
in the thermodynamic limit and, therefore, enable more accurate comparisons with 
experiments, while also helping us gauge finite-size effects in the exact 
diagonalization calculations. Using this method, we compare the experimental 
spin susceptibility from magnetization measurements with 
the numerical results and find very good agreement in a wide range of temperatures. 
Using ED, we also examine models (i) and (ii) and find that results from (i) 
are inconsistent with experimental data for the susceptibility.

Furthermore, we apply the NLCE method to a more general anisotropic-checkerboard-lattice 
Heisenberg model, and tune the ratio of certain exchange constants to capture the 
evolution of thermodynamic quantities in a transition from the planar pyrochlore lattice
to the kagome lattice. These results provide further insight on the 
nature of the spin interactions in the clinoatacamite material and on the effect of
frustration in the kagome and pyrochlore lattices.

The paper is organized as follows: In Sec.~\ref{sec:model}, we introduce the
different models utilized to describe the clinoatacamite compound. Section 
\ref{sec:3d} presents the pyrochlore lattice and its very anisotropic, 
quasi-two-dimensional version, which we use to model clinoatacamite. 
Section \ref{sec:2d} is devoted to the two-dimensional model used. We show how it can 
be seen as a Heisenberg model on an anisotropic checkerboard lattice, and discuss 
its relationship to the uniform kagome lattice and the planar pyrochlore lattice. 
We also describe how NLCEs can be generalized to solve quantum lattice models with 
reduced symmetries, and in particular to solve our two-dimensional model for 
clinoatacamite. In Sec.~\ref{sec:results}, we report the uniform susceptibility 
of clinoatacamite as measured experimentally and our numerical results for the uniform 
susceptibility, specific heat, and entropy obtained within the different theoretical 
models by means of ED and/or NLCE. Finally, our results are summarized in 
Sec.\ \ref{sec:conclusions}.

%%%%%%%%%%%%%%%%%%%%%%%%%%%%%%%%%%%%%%%%%%%%%%%%%%%%%%%%%%%%%%%%%%%%%%%%%%%%%%%%%
%                                    Model                                      %
%%%%%%%%%%%%%%%%%%%%%%%%%%%%%%%%%%%%%%%%%%%%%%%%%%%%%%%%%%%%%%%%%%%%%%%%%%%%%%%%%

\section{Approximate models for clinoatacamite}\label{sec:model}

\subsection{The isotropic and quasi-two-dimensional pyrochlore lattices}
\label{sec:3d}

Clinoatacamite contains three crystallographically distinct Cu sites, such that
the crystal structure consists of kagome planes of Cu2 and Cu3 sites
alternating with triangular planes of Cu1 sites.~\cite{X_Zheng_05a} These sites
are distinguished primarily through the Cu-O-Cu bond angle, with an average
angle of about 96$^{\circ}$ for bonds involving a Cu1 site and an average angle
of about 118$^{\circ}$ for bonds within the Cu2-Cu3 distorted kagome plane. 
(While the distorted lattice structure leads to some further variation within
these averages, the differences are small compared to the difference in
average angles for the in-plane and between-plane cases.) On the basis of these 
differences, it has been suggested that clinoatacamite should be thought of as 
a very-anisotropic pyrochlore (quasi-2D)-lattice Heisenberg model with antiferromagnetic 
kagome planes weakly coupled to triangular planes .~\cite{S_Lee_07}
Within this scenario, and based on bond angle considerations, the exchange 
interaction between layers is likely ferromagnetic and about one order of 
magnitude smaller than the antiferromagnetic in-plane one. 

\begin{figure}[!ht]
{\includegraphics*[width=1.5in]{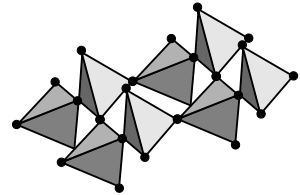}}
\caption{The $16$-site periodic cluster of the pyrochlore lattice.}
\label{fig:3dclusters}
\end{figure}

Other works have emphasized the $\mu_{3}$-OH bridging geometry of clinoatacamite, 
and suggested that the material is best thought of as a distorted pyrochlore 
magnet with exchange interactions that are comparable in the kagome planes
as well as between the kagome and triangular planes.~\cite{X_Zheng_05a,A_Wills_09} 
In Fig.~\ref{fig:3dclusters}, we show the 16-site periodic cluster of the pyrochlore 
lattice that we will use in the ED.

\subsection{The two-dimensional model}\label{sec:2d}

\begin{figure}[!ht]
{\includegraphics*[width=1.5in]{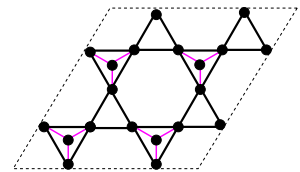}}
\caption{(Color online) The 16-site periodic cluster of the kagome lattice with extra sites inside
down triangles. Pink (thin) bonds represent the coupling between the kagome layer
and the sites sitting on a triangular layer in a 2D model for clinoatacamite.}
\label{fig:2dclusters}
\end{figure}

The study of the thermodynamic properties of the 3D systems in Sec.~\ref{sec:3d} in the thermodynamic 
limit is very demanding using linked-cluster expansions. Hence, we will also model
this material using a two-dimensional geometry consisting of a two-layer system 
of kagome and triangular planes, as depicted in Fig.~\ref{fig:2dclusters}. 
For such a model, we can straightforwardly implement a numerical linked-cluster 
expansion, as explained below. We will show that this simple approximation leads to a very good 
agreement between the experimental uniform susceptibility and the theoretical results.

\begin{figure}[!t]
\centerline {\includegraphics*[width=3.3in]{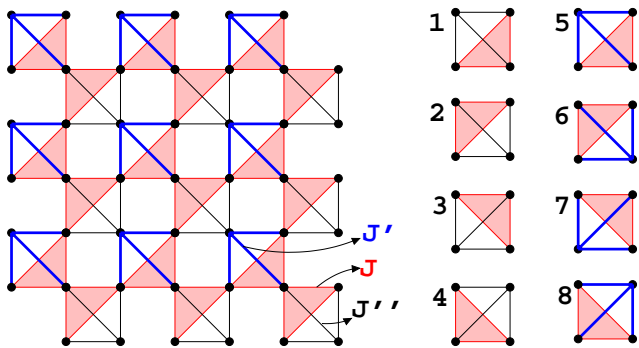}}
\caption{(Color online) The anisotropic checkerboard lattice (left) and the eight realizations of the 
building block used in the square expansion NLCE (right). The shaded area represents 
the kagome lattice in the limit where the red bonds (sides of the shaded triangles) 
have the same strength, $J$, and all other bonds are zero. If the strength of the blue 
(thick) bonds, $J'$, is nonzero, and the interaction on the black (thin) bonds, $J''$, 
is zero, then the resulting structure can represent a kagome lattice coupled to sites 
sitting on a triangular lattice.}
\label{fig:latt}
\end{figure}

In order to perform a NLCE study of such a two-dimensional model, we start with the 
Heisenberg Hamiltonian on the checkerboard lattice,
\begin{equation}
\label{eq:H}
H=\sum_{i,j} J_{ij}\,\hat{{\bf S}}_i\cdot\hat{{\bf S}}_j,
\end{equation}
where $\hat{{\bf S}}_i$ is the spin-$\frac{1}{2}$ vector at site $i$, and $J_{ij}$ 
is the strength of the exchange interaction on each bond that connects sites $i$
and $j$. Throughout the paper, the largest exchange interaction in each case
study sets the unit of energy.
We consider three different types of bonds on the lattice, as seen in
Fig.~\ref{fig:latt}. 
There, the red (shaded) areas make apparent the presence of an embedded kagome 
lattice in the checkerboard lattice. One can immediately see that by tuning the strength 
of the blue (thick) bonds, $J'$, and black (thin) bonds, $J''$, to zero, one captures 
a kagome lattice plus extra decoupled sites. Moreover, if we set $J''$ to zero 
and choose $J'$ ($\neq J$) to be nonzero, then the structure will be that of the kagome 
lattice coupled to sites sitting on a triangular lattice, as depicted in 
Fig.~\ref{fig:2dclusters}. Finally, if $J=J'=J''$, then one has the planar pyrochlore 
lattice. Because of the anisotropies in the Hamiltonian of Eq.~(\ref{eq:H}),
the usual NLCEs for the isotropic case~\cite{E_khatami_11} cannot be used here. Therefore, 
in the following, we implement a NLCE that properly deals with the model 
presented here, in which some of the symmetries of the lattice are broken.

%%%%%%%%%%%%%%%%%%%%%%%%%%%%%%%%%%%%%%%%%%%%%%%%%%%%%%%%%%%%%%%%%%%%%%%%%%%%%%%%%
%                                    Method                                     %
%%%%%%%%%%%%%%%%%%%%%%%%%%%%%%%%%%%%%%%%%%%%%%%%%%%%%%%%%%%%%%%%%%%%%%%%%%%%%%%%%

\subsubsection*{The numerical linked cluster expansion}
\label{sec:formalism}

In linked-cluster expansions,~\cite{linked} an extensive property of the model per 
lattice site in the thermodynamic limit ($P$) is expressed in terms of contributions from 
all of the clusters, up to a certain size, that can be embedded in the lattice:
\begin{equation}
P=\sum_c L(c)w_p(c).
\label{eq:1}
\end{equation}
The contribution from each cluster ($c$) in Eq.~(\ref{eq:1}) is proportional to the weight
of the cluster for that property ($w_p$), and to its multiplicity ($L$). The weight is 
defined recursively as the property for each cluster ($\mathcal{P}$), minus the weights 
of all of its subclusters, 
\begin{equation}
\label{eq:2}
w_p(c)=\mathcal{P}(c)-\sum_{s\subset c}w_p(s),
\end{equation}
and the multiplicity is defined 
as the number of ways that particular cluster can be embedded in the infinite lattice, 
per site. Symmetries of the lattice are often used in identifying topologically 
distinct clusters and in computing their multiplicities. This results in major 
simplifications of the algorithm and usually allows for access to larger clusters 
in the series. Here, we implement NLCEs, where $\mathcal{P}(c)$ is computed by means
of full exact diagonalization,\cite{M_rigol_06,M_rigol_07a} for lattice models that 
break some of the point-group and/or translational symmetries of the underlying lattice. 
In what follows, we discuss how essentially the same expansion as for the symmetric 
case can be used for the latter cases.

As an example, let us consider the uniform checkerboard lattice. In the first order of 
the square expansion, a single crossed square has a multiplicity of 
$1/2$ (Refs.~\onlinecite{M_rigol_07a},\onlinecite{E_khatami_11}) since the number of ways it can be embedded in
the lattice is half the number of sites. In the second order, the only 
distinct cluster consists of two corner-sharing crossed squares. This cluster has a 
multiplicity of $2\times 1/2$, where the extra factor of two comes from the two possibilities 
for its orientation on the lattice (related by a $90^{\circ}$ rotation), and so 
on.\cite{E_khatami_11} Now, consider the anisotropic lattice of Fig.~\ref{fig:latt} where, 
in general, $J''\neq J'\neq J$. In this case, the translational symmetries are reduced by 
a factor of two, and the point-group symmetries are reduced by a factor 
of four, from those of the isotropic checkerboard lattice. So, the square expansion basis 
used for the isotropic case cannot be used for this lattice anymore, since the topological 
clusters and the multiplicities have changed.

The goal is to rearrange the terms in the series to be able
to use the square expansion basis of the isotropic lattice without having to redefine
the topological clusters and their subclusters. Examining the problem more carefully reveals
that the new lattice can still be tiled by considering two different building blocks, as
opposed to one crossed square for the uniform lattice, which is a direct consequence of the
factor-of-two reduction in translational symmetries. These two blocks are numbered $2$ and $5$ 
in Fig.~\ref{fig:latt}. So, in the first order, one has two distinct clusters in the sum, 
each with a multiplicity that is half of that of the single block in the first order of the 
isotropic case. This trend continues in higher orders as for example, in the second order, 
there will be four distinct clusters, as opposed to one in the isotropic case, with 
subclusters that are the two blocks in the first order. But, just like in the first order, 
the multiplicities for each cluster are reduced by a factor proportional to the increasing 
factor in the number of clusters (four for the second order). Moreover, the pool of subclusters
of these four clusters contains the same number of clusters of each type in the first order,
namely, four from each of the two building blocks.

The above argument implies that in the expansion for the less symmetric checkerboard lattice, 
we will have different {\em realizations} of clusters that existed in the expansion for
the symmetric lattice, and that the latter expansion is applicable to the anisotropic case
if the weight of each cluster is replaced by the average weight of those realizations.
It is easy to see that the maximum number of topologically distinct realizations of clusters in the isotropic
square expansion for the lattice of Fig.~\ref{fig:latt} will be eight. 
This number is the same factor by which the point-group and translational 
symmetries are reduced from that of the isotropic checkerboard lattice. In
Fig.~\ref{fig:latt}, we have generated the eight realizations in the first 
order (among which only two are topologically distinct). Each of these building blocks 
can serve as the starting block in the same algorithm that generates all of the clusters in 
the expansion for the isotropic case. In fact, this guarantees the generation of the eight 
realizations for every cluster in the expansion.

The applications of this averaging scheme in NLCEs are not limited to the example 
described here. In principle, this method can be used in any other expansion (e.g., 
site expansion, triangle expansion, etc.), and for any other model with a Hamiltonian
that breaks some symmetries of the underlying lattice. 
In the following section, we use this implementation of the NLCE method to calculate 
the properties of the lattice in Fig.~\ref{fig:latt} for values of the exchange constant 
that transform its symmetry from a uniform planar pyrochlore lattice to near a kagome 
lattice, believed to be the appropriate model for the clinoatacamite compound.

%%%%%%%%%%%%%%%%%%%%%%%%%%%%%%%%%%%%%%%%%%%%%%%%%%%%%%%%%%%%%%%%%%%%%%%%%%%%%%%%%
%                                Case Studies                                   %
%%%%%%%%%%%%%%%%%%%%%%%%%%%%%%%%%%%%%%%%%%%%%%%%%%%%%%%%%%%%%%%%%%%%%%%%%%%%%%%%%

\section{Results}
\label{sec:results}

%=======================================================================

\subsection{Thermodynamics of clinoatacamite}  

\begin{figure}[!t]
\centerline {\includegraphics*[width=3.35in]{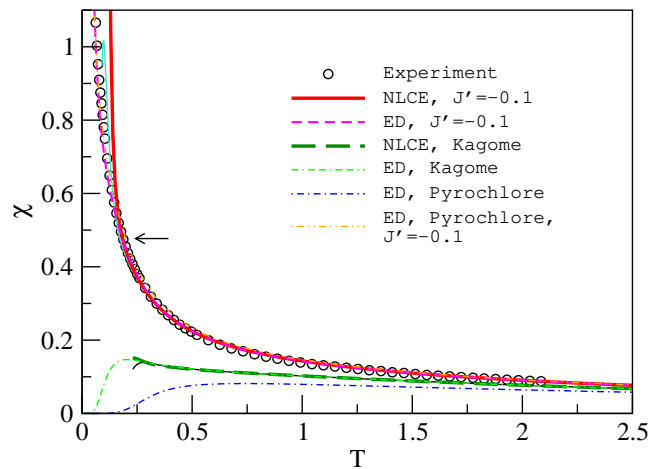}}
\caption{(Color online) Uniform susceptibility per site for clinoatacamite. 
The empty circles are the experimental results. The thick solid line is the 
last order of the NLCE for the system of Fig.~\ref{fig:latt} with $J'=-0.1$ and 
$J''=0$ after the Wynn sum with two cycles of improvement.~\cite{M_rigol_07a} 
The thick dashed line shows the NLCE results for the triangular expansion of the 
kagome-lattice Heisenberg model (from Ref.~\onlinecite{M_rigol_06}).
Thin solid lines are the next-to-last orders of the NLCE sums. In the ED for the 
kagome lattice, we use a $12$-site cluster. For the case with finite $J'$, we use the 
corresponding $16$-site cluster shown in Fig.~\ref{fig:2dclusters}. For the 
pyrochlore lattice, we use the $16$-site cluster shown in Fig.~\ref{fig:3dclusters}.
The arrow marks approximately the point where results from the last two orders of 
NLCE start deviating from each other.}
\label{fig:exp}
\end{figure}

We calculate the thermodynamic properties, such as the specific heat, entropy, and uniform
spin susceptibility for the Hamiltonian (\ref{eq:H}) on the lattice in Fig.~\ref{fig:latt}, 
when $J'=-0.1J$ and $J''=0$, to represent clinoatacamite. 
A 16-site periodic cluster of the resulting lattice is depicted in Fig.~\ref{fig:2dclusters}
with thick (thin) bonds representing $J$ 
($J'$). Note, however, that NLCE computes these properties directly for the infinite system 
and does not have any statistical or systematic errors (such as finite-size effects) within 
its region of convergence in temperature. We carry out the NLCE calculations to the sixth order (six building 
blocks with maximum $19$ sites) of the square expansion. 

In Fig.~\ref{fig:exp}, we show the 
spin susceptibility per site from the last two orders of NLCE for this system. There, we have
also included the experimental data for this material. 
The magnetic susceptibility of a polycrystalline clinoatacamite sample was
measured with a SQUID magnetometer under an applied field of 500 Oe.  The
susceptibility was measured while warming from 2 to 400 K after field cooling.
Consistent with previously published susceptibility results,~\cite{X_Zheng_05a}
a weak ferromagnetic moment is 
observed below $T_{c1}$ $\approx$ 6.4 K (not shown) and a subtle kink is observed in the
susceptibility at $T_{c2}$ = 18 K.  We will focus on the susceptibility above 10
K, where the experimental data can be compared with the numerical results.
The experimental molar susceptibility in cgs units is related to the numerical
one by $\chi_{\textrm{exp}}=C\chi$, where the constant $C=N_Ag^2\mu_{B}^2/kJ =
0.3752g^2/J$. We use $J=193$ K from the Curie-Weiss formula, and take 
$g=2.14$ so that the numerical and experimental susceptibilities match at the
highest 
temperature available experimentally ($T\sim 2.1$). There is a remarkable agreement between 
the experiment and this approximate model for all of the temperatures above the convergence 
temperature of the NLCE ($\sim 0.2$, indicated by the arrow in Fig.~\ref{fig:exp}). To have a 
better idea about the effect of the extra sites of the triangular layer on the susceptibility 
of the kagome lattice, we also show results from a triangle-based NLCE on the kagome 
lattice with up to eight triangles.\cite{M_rigol_06,M_rigol_07a} 

\begin{figure}[!t]
\centerline {\includegraphics*[width=3.35in]{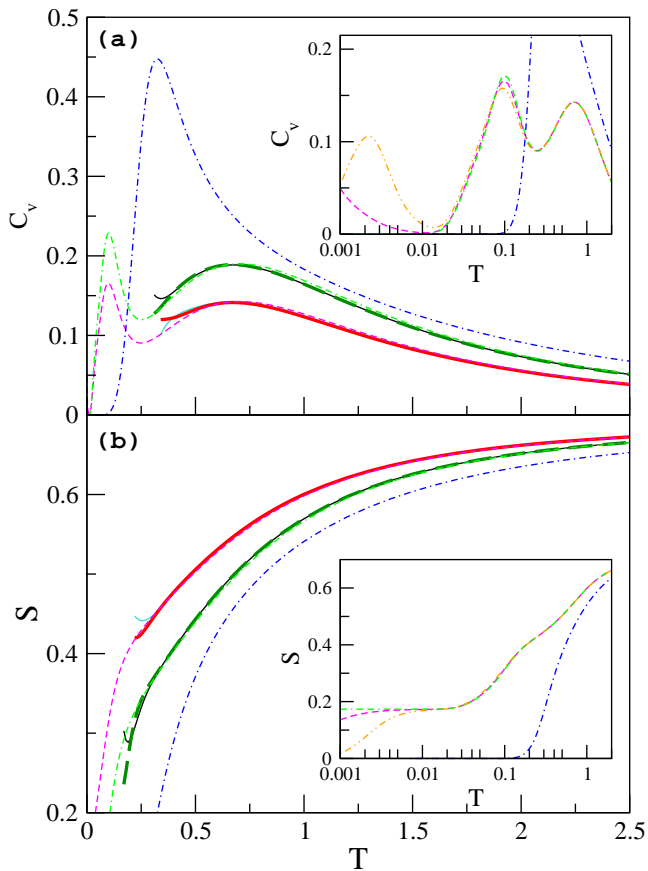}}
\caption{(Color online) NLCE and ED results for (a) the specific heat and (b) the entropy per site of the 
Heisenberg model for clinoatacamite as well as on the kagome and pyrochlore lattices.
Lines are the same as in Fig.~\ref{fig:exp}. The inset of (a) shows the ED results on a
logarithmic temperature grid. We have multiplied the kagome-lattice results by 
$\frac{3}{4}$ to establish a fair comparison with the results for the lattice of 
Fig.~\ref{fig:2dclusters}, and have included those for the $16$-site pyrochlore-lattice Heisenberg 
model, where the coupling between kagome layers is set to $J'=-0.1$. The inset in (b)
is the same as the inset in (a) for the entropy, except that the entropy of an isolated
spin is also properly added to that of the kagome lattice (see text).}
\label{fig:theor}
\end{figure}

It is clear that the extra sites with weak ferromagnetic couplings are responsible for the 
enhancement of the uniform susceptibility at low temperatures. To understand this, we consider 
the limiting case where the sites on the triangular layer are completely decoupled from the 
ones on the kagome layer ($J'=0$). In the thermodynamic limit, since the kagome layer 
contains only $3/4$ of the sites, any property per site can be written 
as $P=\frac{1}{4}\mathcal{P}_0+\frac{3}{4}\mathcal{P}_{\text{kgm}}$, where $\mathcal{P}_0$ 
is the property for a single site and $\mathcal{P}_{\text{kgm}}$ is the property per site for 
the kagome lattice. Therefore, in the case of susceptibility, a zero-temperature divergence 
will emerge from the susceptibility of an isolated spin, $\chi_0=\frac{1}{4T}$. In fact, if we 
take $\chi_{\text{kgm}}$ to be the NLCE results for the kagome lattice and calculate 
$\chi=\frac{1}{4}\chi_0+\frac{3}{4}\chi_{\text{kgm}}$, then the resulting curve lies very close,
but slightly below, that of the NLCE with $J'=-0.1$ (see Fig.~\ref{fig:SUS}), i.e., the 
divergence in the uniform susceptibility of clinoatacamite is mostly due to the nearly isolated 
interlayer spins. However, a small negative $J'$ presumably produces a finite temperature 
ordering transition in the three-dimensional material, which is observed in the experiments 
at $\sim 6$ K.

The results from ED on finite clusters with periodic boundary condition further support these 
findings. In Fig.~\ref{fig:exp}, we show the spin susceptibility for the $16$-site cluster 
of Fig.~\ref{fig:2dclusters}, and the quasi-2D model, with $J'=-0.1$. They both agree 
with the experimental results very well in the entire temperature range. We also show 
the ED results for the corresponding $12$-site cluster on the kagome 
lattice (which is the same cluster as in Fig.~\ref{fig:2dclusters}, but without the extra sites inside 
down triangles) and the uniform pyrochlore lattice of Fig.~\ref{fig:3dclusters}. The latter 
largely disagrees with the experimental results, invalidating the proposals that clinoatacamite 
has such uniformity in exchange constants.~\cite{X_Zheng_05a,A_Wills_09}

At this time, the lack of a nonmagnetic isostructural compound has made it impossible to
accurately determine the lattice contribution to the specific heat over the
temperature range where NLCEs are valid. Therefore, we cannot currently compare
the magnetic specific heat of clinoatacamite with the results of numerical 
calculations the way we have with the susceptibility. Nevertheless, in Fig.~\ref{fig:theor}, 
we show the numerical results for the entropy and the specific heat for the models of 
clinoatacamite and the other systems discussed above, which could be use to compare with future 
experiments. Since the specific heat for an isolated spin is zero, the values for the 
$J'=-0.1$ case in Fig.~\ref{fig:theor}(a) are roughly $\frac{3}{4}$ of those for the kagome 
lattice, at least for $T\gtrsim |J'|$ [see also the inset of Fig.~\ref{fig:theor}(a)]. 
The position of one of the peaks, captured in the ED calculations both for
the pure kagom{\'e} and the model for clinoatacamite at $T\sim 0.1$,
approximately coincides with the $18$ K peak observed in the experiments,
considering $J\sim 193$ K.~\cite{X_Zheng_05a,X_Zheng_05b,H_Morodomi_10} The
existence of such a peak in the specific heat of the kagome-lattice
Heisenberg model has been a topic of discussion for a long time,
\cite{v_elser_89,n_elstner_94,g_misguich_05,M_rigol_07a} and the experiments
with the clinoatacamite compound may have provided a proof of its existence.
On the other hand, the only 
peak of the specific heat for the finite-size pyrochlore lattice from ED is at $T\sim 0.3$. 
This is inconsistent with the experimental results for clinoatacamite and is yet another 
evidence that this material is not well described by the uniform (or nearly uniform) 
pyrochlore Heisenberg model. 

In the inset of Fig.~\ref{fig:theor}(a), we show the specific heat from ED on a logarithmic 
temperature scale and down to $T=0.001$. The specific heat of the kagome or the pyrochlore 
lattice vanishes below $T\sim 0.01$, whereas a third peak emerges at $T<0.001$ for the 2D 
model of clinoatacamite (the cluster of Fig.~\ref{fig:2dclusters}). A similar 
feature also exists in the corresponding quasi-2D model with $J'=-0.1$. 
The peak moves to higher temperatures
by increasing $|J'|$. Although finite-size effects often prevent ED from predicting, even 
qualitatively, the correct features of such models with long-range order at low temperatures, 
the appearance of this low-temperature peak due to the finite $J'$ may signal a possible 
very-low-temperature phase transition in the thermodynamic limit, 
perhaps associated with the one observed experimentally for clinoatacamite at $T\sim6$ K.

The entropies per site for the 2D and quasi-2D models of clinoatacamite, the 
kagome-lattice and the pyrochlore-lattice Heisenberg models are shown in 
Fig.~\ref{fig:theor}(b). Just like for the specific heat, we show, in the inset of 
Fig.~\ref{fig:theor}(b), the low-temperature entropy of different models from the ED, 
which give us an idea of what may happen at lower temperatures. There, 
we have multiplied the entropy of the $12$-site kagome lattice by $\frac{3}{4}$ 
and added the contribution from the isolated spins ($\frac{\ln2}{4}$) to be able to
properly compare it to the entropy of the 16-site clusters. We note that above 
$T=0.01$, all entropies but the one for the pyrochlore lattice agree with each other. 
Also, as inferred from the specific-heat plots, a finite $J'$ seems to bring about 
a phase transition at a very low temperature, after which the entropy drops to zero.

\subsection{Transition between planar pyrochlore and kagome lattices}

\begin{figure}[!t]
\centerline {\includegraphics*[width=0.47\textwidth]{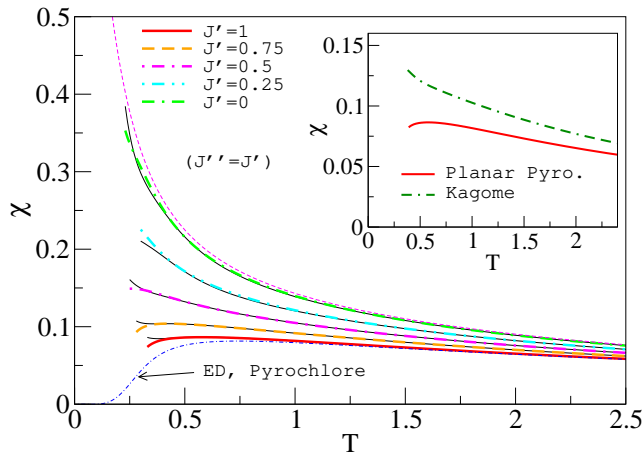}}
\caption{(Color online) NLCE results for the uniform susceptibility per site of the Heisenberg model 
in the transition between the planar pyrochlore lattice ($J''=J'=J$ in the lattice of 
Fig.~\ref{fig:latt}) and the kagome lattice with extra decoupled sites ($J''=J'=0$). 
The thin dashed (dotted-dashed) line is the ED result for the 2D model of clinoatacamite
(uniform pyrochlore lattice). For $J'=0$ 
and $0.25$, black (thin) solid lines and color (thick) lines are the fifth and sixth 
orders of the bare sums in the expansion, respectively. For all other values of $J'$, 
we have used Wynn extrapolation with one cycle of improvement,~\cite{M_rigol_07a} for 
which the thin solid and thick lines are the last two orders. The inset compares the 
uniform susceptibility per site for the planar pyrochlore lattice and the kagome lattice.}
\label{fig:SUS}
\end{figure}

To gain further insights about how thermodynamic properties change in transitions 
between different frustrated models, and its implications for the research on future materials,
we study here the uniform susceptibility, and the specific heat in the transition between 
the planar pyrochlore lattice and the kagome lattice, using the implementation of NLCE 
described in Sec.~\ref{sec:formalism}. We start with the former lattice ($J''=J'=J$). 
To approach the kagome lattice, we simultaneously decrease $J'$ and $J''$ 
from $1$ to $0$. As discussed above, the latter limit represents the kagome-lattice 
Heisenberg model plus an extra isolated spin for every three spins in the kagome 
lattice, which is closely related to the 2D model for clinoatacamite.

As can be seen in Fig.~\ref{fig:SUS}, the spin susceptibility of the planar pyrochlore 
lattice can even provide a good estimate for that of the 3D pyrochlore lattice (from ED), as 
the difference between the two remains relatively small for temperatures accessible to NLCE 
($T>0.3$). To show the proximity of the results on the other side of the transition to 
the model for clinoatacamite ($J'=-0.1$ and $J''=0$), we plot in Fig.~\ref{fig:SUS} 
results from the latter from ED. As the spins on the triangular layer decouple from those 
on the kagome layer by decreasing $J'$, the $\frac{1}{T}$ divergent signature of the 
susceptibility of isolated spins, similar to what has been seen in the experiments on 
clinoatacamite, becomes apparent.

It is now interesting to compare the uniform susceptibility for the planar pyrochlore lattice 
and the pure kagome lattice. Within the present NLCE calculation, the latter can be 
obtained by subtracting the contribution of isolated spins in the $J''=J'=0$ case.
The results are shown in the inset of Fig.~\ref{fig:SUS}. One can clearly see there that 
the kagome lattice has a higher uniform susceptibility than the planar pyrochlore lattice 
for all temperatures accessible within our NLCE.

\begin{figure}[!t]
\centerline {\includegraphics*[width=0.47\textwidth]{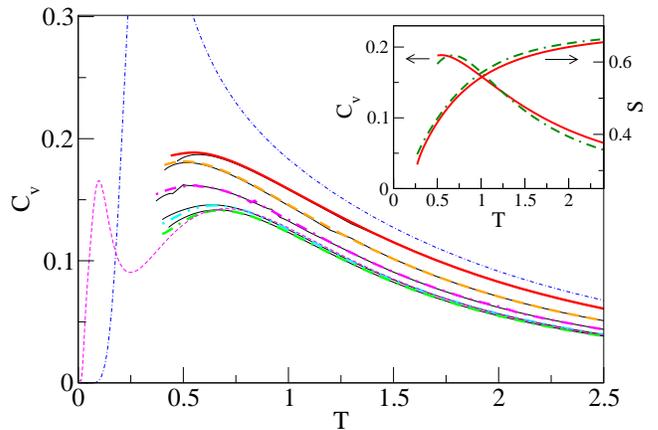}}
\caption{(Color online) NLCE results for the specific heat and entropy per site of the Heisenberg model on 
the anisotropic checkerboard lattice of Fig.~\ref{fig:latt}, with $0\le J''=J'\le J$.
The inset compares the specific heat and the entropy per site of the planar pyrochlore 
and the kagome lattices. The lines are the same as in Fig.~\ref{fig:SUS}}
\label{fig:SH}
\end{figure}

The planar pyrochlore lattice and the pure kagome lattice are two of the most frustrated
lattices known. In Fig.~\ref{fig:SH}, we show how the specific heat evolves in the transition
between them for the same parameters depicted in Fig.~\ref{fig:SUS}.
Unlike the spin susceptibilities, the specific heat of the planar pyrochlore lattice is
qualitatively different from the pyrochlore lattice. In the two-dimensional model, as $J'$ and $J''$ 
decrease, the high-temperature peak is suppressed. However, this is largely due to the fact that
one-fourth of the spins in the system are decoupled from the lattice in the limit of 
$J''=J'=0$ and, therefore, have vanishing specific heat. Consequently, if one compares the entropy 
and specific heat per site of the planar-pyrochlore-lattice and the kagome-lattice Heisenberg models
(inset in Fig.~\ref{fig:SH}), one sees that their values are in fact very close for all of the 
temperatures calculated here. Interestingly, this shows that both lattices have a very similar 
degree of frustration.

\section{Conclusions}\label{sec:conclusions}

We have presented a numerical study of the thermodynamic properties for 
models of the clinoatacamite compound. In particular, we computed the spin susceptibility, entropy, 
and specific heat, using the ED of finite periodic clusters and an implementation of the NLCEs 
that properly deals with the breaking of lattice symmetries introduced by the particular 
model Hamiltonian of interest. We find an
excellent agreement between the experimental uniform susceptibility of clinoatacamite
from magnetic measurements and our numerical results for the Heisenberg model 
on a lattice that consists of a kagome layer, coupled weakly to a triangular layer.
Together with a study of the entropy and the specific heat of the kagome and 
pyrochlore lattices, we provide strong evidence that clinoatacamite has a pyrochlore 
structure with only weak ferromagnetic coupling between its kagome layers. Employing 
our generalized NLCE, we also studied the above thermodynamic quantities in a transition 
between the planar pyrochlore lattice, which has a uniform susceptibility similar to 
that of the pyrochlore lattice, and the kagome lattice plus isolated spins, closely 
related to the model for the clinoatacamite compound.

\section*{Acknowledgments}
This research was supported by the NSF under Grant No.~OCI-0904597.
We thank Young Lee for guidance with the susceptibility measurements,
and Matthew Shores, Bart Bartlett, Emily Nytko, and Daniel Nocera for
providing the clinoatacamite sample.

\end{document}